\documentclass[twocolumn,english]{revtex4}
\usepackage[T1]{fontenc}
\usepackage[latin9]{inputenc}
\usepackage{color}
\usepackage{babel}
\usepackage{amsmath}
\usepackage{graphicx}
\usepackage[unicode=true]
{hyperref}

\makeatletter

\providecommand{\tabularnewline}{\\}

\@ifundefined{textcolor}{}
{%
 \definecolor{BLACK}{gray}{0}
 \definecolor{WHITE}{gray}{1}
 \definecolor{RED}{rgb}{1,0,0}
 \definecolor{GREEN}{rgb}{0,1,0}
 \definecolor{BLUE}{rgb}{0,0,1}
 \definecolor{CYAN}{cmyk}{1,0,0,0}
 \definecolor{MAGENTA}{cmyk}{0,1,0,0}
 \definecolor{YELLOW}{cmyk}{0,0,1,0}
}

\makeatother

\begin{document}
\title{Chemical Reactivity Studies by the Natural-Orbital-Functional 2nd-Order-Møller-Plesset
(NOF-MP2) method. Water Dehydrogenation by the Scandium Cation}
\author{Jose M. Mercero$^{1}$, Jesus M. Ugalde$^{1}$ and Mario Piris$^{1,2}$\bigskip{}
}
\address{$^{1}$Kimika Fakultatea, Euskal Herriko Unibertsitatea (UPV/EHU)
and Donostia International Physics Center (DIPC), P.K. 1072, 20080
Donostia, Euskadi, Spain. \\
$^{2}$Basque Foundation for Science (IKERBASQUE), 48013 Bilbao, Euskadi,
Spain.\bigskip{}
}
\begin{abstract}
The reliability of the recently proposed natural orbital functional
supplemented with second-order Møller-Plesset calculations, (NOF-MP2),
has been assessed for the mechanistic studies of elementary reactions
of transition metal compounds by investigating the dehydrogenation
of water by the scandium cation. Both high- and low-spin state potential
energy surfaces have been searched thoroughly. Special attention has
been paid to the assessment of the capability of the NOF-MP2 method
to describe the strong, both static and dynamic, electron correlation
effects on the reactivity of Sc$^{+}$($^{3}$D, $^{1}$D) with water.
In agreement with experimental observations, our calculations correctly
predict that the only exothermic products are the lowest-lying ScO$^{+}(^{1}\mathrm{\Sigma})$
and H$_{2}(^{1}\mathrm{\Sigma}_{g}^{+})$ species. Nevertheless, an
in-depth analysis of the reaction paths leading to several additional
products was carried out, including the characterization of various
minima and several key transition states. Our results have been compared
with the highly accurate multiconfigurational supplemented with quasi
degenerate perturbation theory, MCQDPT, wavefunction-type calculations,
and with the available experimental data. It is observed that NOF-MP2
is able to give a satisfactorily quantitative agreement, with a performance
on par with that of the MCQDPT method. 

\bigskip{}

\bigskip{}
\end{abstract}
\maketitle

\section{Introduction}

Advanced quantum mechanical methods for the determination of the molecular
electronic structure have, in the last decade, evolved towards reliable
computational algorithms for the elucidation of the mechanisms of
chemical reactions, both in the condensed and in the gas phase \cite{science:2015:catalysis}.
Gas phase studies, in particular, constitute the ideal playground
where theory and experiment converge straightforwardly. The strictly
molecular nature of both approaches, enables direct comparison of
their results free from ``environmental'' perturbations. The large
and increasing number of combined theoretical/experimental chemical
reaction mechanistic studies, indisputably shows that the development
of reliable and computationally efficient density functional theory
(DFT) based methods has contributed greatly to this present status
\cite{koch:2001}.

In this vein, the reactions of transition metal cations with a large
variety of substrates, including both first- and second-row hydrides
\cite{o2,o3,o7,o12,o15,oier2013} and, in particular water \cite{Clemmer1993,aran1999,Irigoras1999,Irigoras2000,aran2004},
have been exhaustively studied. Thus, the studies alluded to above
have firmly established that the earlier transition metal cations
(Sc$^{+}$, Ti$^{+}$, V$^{+}$, Cr$^{+}$ and Mn$^{+}$) are more
reactive than their corresponding oxides, while the opposite is true
for the late transition metal cations (Co$^{+}$, Ni$^{+}$ and Cu$^{+}$).

The earlier transition metal cations have high-spin ground states,
while the ground states of their corresponding oxide cations are low-spin.
For late transition metals the opposite is found, namely, the metal
cations have low-spin ground states and their corresponding oxides
have high-spin ground states. This precludes complex reaction mechanisms
for which allowance for spin-crossings to occur should be made \cite{shaik:2020}.
Iron, lies in the middle ground, i.e.: Fe$^{+}$ and FeO$^{+}$ have
both high-spin ground states, a fact that does not prevent them from
spin-crossings to occur \cite{Irigoras1999}.

The distinct gas-phase reactivity of the transition metal cations
can naturally be ascribed to the electronic configuration and to the
spin state \cite{o1} of the reacting metal's cation. However, the
electronic structure of compounds having transition metals are tough
for most approximate exchange-correlation functional based DFT methods,
for they lack proper consideration of the strong dynamical and static
correlation effects arising from the incomplete $d$-shells of the
metals \cite{txema:2005}. A fact which forces theoretical analyses
to be performed at a higher level of accuracy, involving in many cases
multiconfigurational wave-function type methods to properly account
for the electron correlation effects. Needless to say that this increases
enormously the computational effort and, unfortunately, more often
than not it ends up in unbearable computational demands.

Earlier in the 80's, it was suggested that natural orbital functional
theory (NOFT) implementations could be an attractive alternative formalism
to current wave-function based algorithms. In spite of the fact that
computational schemes based on exact functionals \cite{Gilbert1975,Levy1979,Valone1980}
were found to be too expensive from a computational point of view,
approximate, albeit rigorous, natural orbital functionals (NOFs) have
been developed for practical applications. Approximate NOFs have demonstrated
\cite{Mitxelena2019} to be more accurate than electron density functionals
for highly multiconfigurational systems in particular, and to scale
more satisfactorily than multiconfigurational wave-function type methods
with respect to the number of basis functions. An extensive account
of the formulation and the development of such approximate NOFs can
be found elsewhere \cite{Piris2007,Piris2014a,Pernal2016,Piris2018a,Piris2018b}.

Recently, an open-source implementation of NOF based methods has been
made available to the quantum chemistry community \cite{Piris2021}.
The associated computer program \textcolor{cyan}{\href{http://github.com/DoNOF/DoNOFsw}{\underline{DoNOF}}}
is designed to solve the energy minimization problem of an approximate
NOF, which describes the ground-state of an N-electron system in terms
of the natural orbitals (NOs) and their occupation numbers (ONs).
The program includes the NOFs developed in the Donostia quantum chemistry
group. In this paper, we use the Piris NOF 7 (PNOF7) approximate NOF,
which has proven \cite{Mitxelena2020,Mitxelena2020a} to be an efficient
and accurate enough alternative for strongly correlated electrons.

It is worth noting that PNOF7 is able to describe the complete intra-pair,
but only the static inter-pair electron correlation. In order to recover
the missing inter-pair dynamic electron correlation, a single-reference
global method for the electron correlation was introduced \cite{Piris2017,Piris2019},
taking as a reference the Slater determinant formed with the NOs of
an approximate NOF, in our case PNOF7.

Within this approach, denoted as natural orbital functional - second-order
Møller-Plesset (NOF-MP2) method, the total energy of an N-electron
system can be cast as, 
\begin{equation}
E=\tilde{E}_{hf}+E^{corr}=\tilde{E}_{hf}+E^{dyn}+E^{sta}\label{Etotal}
\end{equation}
where $\tilde{E}_{hf}$ is the Hartree-Fock (HF) energy obtained with
the NOs, the dynamic energy ($E^{dyn}$) is derived from a modified
MP2 perturbation theory, while the non-dynamic energy ($E^{sta}$)
is obtained from the static component of the employed NOF. Actually,
the correction $E^{dyn}$ is based on an orbital-invariant formulation
of the MP2 energy \cite{Piris2018c}. It has been observed that NOF-MP2
is able to give a quantitative agreement for dissociation energies,
with a performance comparable to that of the accurate CASPT2 method
in hydrogen abstraction reactions \cite{Lopez2019}.

In the present paper, we analyze and assess the performance of NOF-MP2
for the description of the Sc$^{+}$ + H$_{2}$O $\rightarrow$ ScO$^{+}$
+ H$_{2}$ reaction. This is the first time that NOF based methods
have been used for chemical reaction mechanistic studies of transition
metal containing compounds.

We address the singlet and triplet spin-states, with the NOFT for
multiplets formulation \cite{Piris2019}. The results of our NOF-MP2
(over PNOF7 geometries) calculations are compared with the energies
obtained by the MCQDPT/MCSCF(10,17) level of theory (hereafter MCQDPT),
as implemented in GAMESS-US \cite{gamess} program package. The TZPV+
basis set, which has been found to give excellent results for transition
metal cation reactions with water \cite{aran1999,Irigoras2000,Irigoras1999},
has been used for all the calculations. This basis set is built starting
from the TZVP basis developed by Alrichs et. al. \cite{alrichs1,alrichs2},
complemented by two sets of $p$ \cite{hay}, one set of $d$ \cite{Watchers}
and three uncontracted $f$ functions \cite{raghavachari}. The results
shown below reconfirm the reliability of this basis set. Frequencies
where calculated for all the reported stationary points and ZPVE corrections
were applied to the reported energies. The article is organized as
follows. In section 2, we briefly lay the foundations of the NOFT.
In section 3, we assess the quality of the results obtained, and finally,
in section 4 the conclusions are presented.

\section{Theory}

The electronic energy of an approximate NOF is given in terms of the
NOs $\left\{ \phi_{i}\right\} $ and their ONs $\left\{ n_{i}\right\} $
as follows 
\begin{equation}
E=\sum\limits _{i}n_{i}\mathcal{H}_{ii}+\sum\limits _{ijkl}D[n_{i},n_{j},n_{k},n_{l}]\left<kl|ij\right>\label{NOF}
\end{equation}
where $\mathcal{H}_{ii}$ denotes the diagonal elements of the one-particle
part of the Hamiltonian involving the kinetic energy and the external
potential operators, $<kl|ij>$ are the matrix elements of the two-particle
interaction, and $D[n_{i},n_{j},n_{k},n_{l}]$ represents the reconstructed
two-particle reduced density matrix (2RDM) from the ONs. Restriction
of the ONs to the range $0\leq n_{i}\leq1$ represents a necessary
and sufficient condition for ensemble $\mathrm{N}$-representability
of the one-particle reduced density matrix (1RDM) \cite{Coleman1963}
under the normalization condition $\sum_{i}n_{i}=\mathrm{N}$.

Our non-relativistic Hamiltonian is free of spin coordinates, hence
a state with total spin $S$ is a multiplet, i.e., a mixed quantum
state (ensemble) that allows all possible $S_{z}$ values. This approach
differs from the methods routinely used in electronic structure calculations
that focus on the high-spin component or break the spin symmetry.
Next, we briefly describe how we do the reconstruction of $D$ to
achieve PNOF7 for spin-multiplets. A more detailed description can
be found in Ref. \cite{Piris2019}.

We consider $\mathrm{N_{I}}$ single electrons which determine the
spin $S$ of the system, and the rest of electrons ($\mathrm{N_{II}}=\mathrm{N-N_{I}}$)
are spin-paired, so that all spins corresponding to $\mathrm{N_{II}}$
electrons provide a zero spin. We focus on the mixed state of highest
multiplicity: $2S+1=\mathrm{N_{I}}+1,\,S=\mathrm{N_{I}}/2$. For example,
for two single electrons, $\mathrm{N_{I}}=2$ and $S=1$, then we
have a mixed state $\left\{ \left|1,-1\right\rangle ,\left|1,0\right\rangle ,\left|1,1\right\rangle \right\} $
which forms a triplet state. In the absence of single electrons ($\mathrm{N_{I}}=0$),
the energy (\ref{NOF}) obviously reduces to a NOF that describes
singlet states.

For an ensemble of pure states $\left\{ \left|SM_{s}\right\rangle \right\} $,
we note that 
\begin{equation}
\langle\hat{S}_{z}\rangle=\frac{1}{\mathrm{N_{I}}+1}{\textstyle {\displaystyle \sum_{M_{s}=-\mathrm{N_{I}}/2}^{\mathrm{N_{I}}/2}}M_{s}}=0.\label{SZ0}
\end{equation}

Eq. (\ref{SZ0}) implies that the expected value of $\hat{S}_{z}$
for the whole ensemble is zero. Consequently, the spin-restricted
theory can be adopted even if the total spin of the system is not
zero as is the case in the triplet state. We use a single set of orbitals
for $\alpha$ and $\beta$ spins. All the spatial orbitals will be
then doubly occupied in the ensemble, so that occupancies for particles
with $\alpha$ and $\beta$ spins are equal: $n_{p}^{\alpha}=n_{p}^{\beta}=n_{p}.$

The next step is dividing the orbital space $\Omega$ into two subspaces:
$\Omega=\Omega_{\mathrm{I}}\oplus\Omega_{\mathrm{II}}$. $\Omega_{\mathrm{II}}$
is composed of $\mathrm{N_{II}}/2$ mutually disjoint subspaces $\Omega{}_{g}$.
Each of which contains one orbital $\left|g\right\rangle $ with $g\leq\mathrm{N_{II}}/2$,
and $\mathrm{N}_{g}$ orbitals $\left|p\right\rangle $ with $p>\mathrm{N_{II}}/2$,
namely, 
\begin{equation}
\Omega{}_{g}=\left\{ \left|g\right\rangle ,\left|p_{1}\right\rangle ,\left|p_{2}\right\rangle ,...,\left|p_{\mathrm{N}_{g}}\right\rangle \right\} .\label{OmegaG}
\end{equation}

Taking into account the spin, the total occupancy for a given subspace
$\Omega{}_{g}$ is 2, which is reflected in the following sum rule:
\begin{equation}
\sum_{p\in\Omega_{\mathrm{II}}}n_{p}=n_{g}+\sum_{i=1}^{\mathrm{N}_{g}}n_{p_{i}}=1,\quad g=1,2,...,\frac{\mathrm{N_{II}}}{2}.\label{sum1}
\end{equation}
In general, $\mathrm{N}_{g}$ may be different for each subspace,
but it should be sufficient for the description of each electron pair.
In this work, $\mathrm{N}_{g}$ is equal to a fixed number for all
subspaces $\Omega{}_{g}\in\Omega_{\mathrm{II}}$. The maximum possible
value of $\mathrm{N}_{g}$ is determined by the basis set used in
calculations. In the current work, our calculations indicated that
$\mathrm{N}_{g}=2$ is sufficient to correctly describe the reaction
of Sc$^{+}$ with water. In fact, we consider only the ONs that do
not exceed a certain threshold, e.g. 0.01. Small ONs are known to
contribute only to the dynamic correlation that will be accounted
for by the MP2 correction to PNOF7. From (\ref{sum1}), it follows
that 
\begin{equation}
2\sum_{p\in\Omega_{\mathrm{II}}}n_{p}=2\sum_{g=1}^{\mathrm{N_{II}}/2}\left(n_{g}+\sum_{i=1}^{\mathrm{N}_{g}}n_{p_{i}}\right)=\mathrm{N_{II}}.\label{sumNpII}
\end{equation}

Here, the notation $p\in\Omega_{\mathrm{II}}$ represents all the
indexes of $\left|p\right\rangle $ orbitals belonging to $\Omega_{\mathrm{II}}$.
It is important to recall that orbitals belonging to each subspace
$\Omega_{g}$ vary along the optimization process until the most favorable
orbital interactions are found. Therefore, orbitals do not remain
fixed in the optimization process, they adapt to the problem.

Similarly, $\Omega_{\mathrm{I}}$ is composed of $\mathrm{N_{I}}$
mutually disjoint subspaces $\Omega{}_{g}$. In contrast to $\Omega_{\mathrm{II}}$,
each subspace $\Omega{}_{g}\in\Omega_{\mathrm{I}}$ contains only
one orbital $g$ with $2n_{g}=1$. It is worth noting that each orbital
is completely occupied individually, but we do not know whether the
electron has $\alpha$ or $\beta$ spin: $n_{g}^{\alpha}=n_{g}^{\beta}=n_{g}=1/2$.
It follows that 
\begin{equation}
2\sum_{p\in\Omega_{\mathrm{I}}}n_{p}=2\sum_{g=\mathrm{N_{II}}/2+1}^{\mathrm{N_{\Omega}}}n_{g}=\mathrm{N_{I}}.\label{sumNpI}
\end{equation}

In Eq. (\ref{sumNpI}), $\mathrm{\mathrm{N}_{\Omega}=}\mathrm{N_{II}}/2+\mathrm{N_{I}}$
denotes the total number of suspaces in $\Omega$. Taking into account
Eqs. (\ref{sumNpII}) and (\ref{sumNpI}), the trace of the 1RDM is
verified equal to the number of electrons: 
\begin{equation}
2\sum_{p\in\Omega}n_{p}=2\sum_{p\in\Omega_{\mathrm{II}}}n_{p}+2\sum_{p\in\Omega_{\mathrm{I}}}n_{p}=\mathrm{N_{II}}+\mathrm{N_{I}}=\mathrm{\mathrm{N}}.\label{norm}
\end{equation}

\begin{figure*}
\begin{centering}
\textcolor{black}{\caption{\label{fig1} Splitting of the orbital space $\Omega$ into subspaces
used in the study of the reaction $Sc\ensuremath{^{+}}+H\ensuremath{_{2}}O\ensuremath{\rightarrow}ScO\ensuremath{^{+}}+H\ensuremath{_{2}}$.
The arrows depict the values of the ensemble occupation numbers, alpha
($\downarrow$) or beta ($\uparrow$), in each orbital. The seven
innermost orbitals with frozen electrons are not shown.\bigskip{}
}
}
\par\end{centering}
\centering{}\textcolor{black}{\includegraphics[scale=0.48]{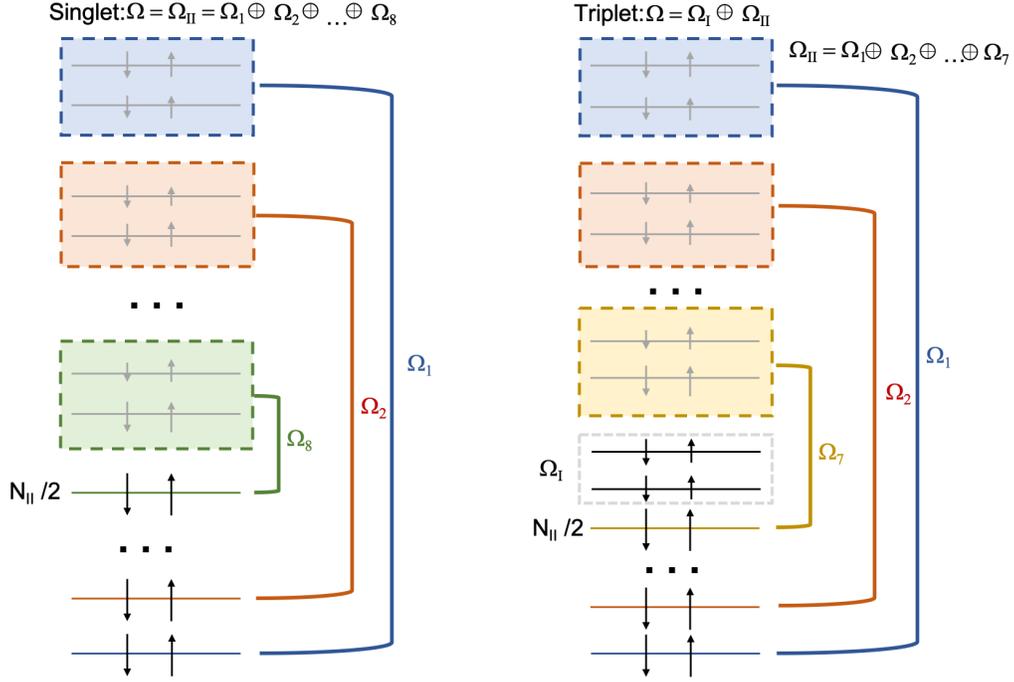}}
\end{figure*}

In Fig. \ref{fig1}, we show the splitting into subspaces of the orbital
space $\Omega$ used in the study of the reaction Sc$^{+}$ + H$_{2}$O
$\rightarrow$ ScO$^{+}$ + H$_{2}$. On the right, the division employed
for the triplet ($S=1$, $\mathrm{N_{I}}=2$) is depicted, where two
orbitals make up the subspace $\Omega_{\mathrm{I}}$, whereas fourteen
electrons ($\mathrm{N_{II}}=14$) distributed in seven subspaces $\left\{ \Omega_{1},\Omega_{2},...,\Omega_{7}\right\} $
make up the subspace $\Omega_{\mathrm{II}}$. On the left is the split
used for the singlet ($S=0$, $\mathrm{N_{I}}=0$), hence sixteen
electrons ($\mathrm{N_{II}}=16$) distributed in eight subspaces $\left\{ \Omega_{1},\Omega_{2},...,\Omega_{8}\right\} $
compose the subspace $\Omega_{\mathrm{II}}$. Note that the value
of $\mathrm{N}_{g}$ has been set equal to two in both cases. Also,
seven pairs of electrons are kept frozen in the innermost orbitals
and are not shown in the figure.

\textcolor{black}{}

The construction \cite{Piris2018a,Piris2010a} of an N-representable
functional given by Eq. (\ref{NOF}) is related to the N-representability
problem of $D$ \cite{Mazziotti2012}. Using its ensemble N-representability
conditions to generate a reconstruction functional leads to PNOF7
\cite{Piris2017,Piris2019,mitxelena2018a}. Assuming real spatial
orbitals, we obtain 
\begin{equation}
E^{PNOF7}=\sum\limits _{g=1}^{\mathrm{N_{II}}/2}E_{g}+\sum\limits _{g=\mathrm{N_{II}}/2+\mathrm{1}}^{\mathrm{N}_{\Omega}}\mathcal{H}_{gg}+\sum\limits _{f\neq g}^{\mathrm{N}_{\Omega}}E_{fg}\label{EPNOF7}
\end{equation}
where 
\begin{equation}
E_{g}=2\sum\limits _{p\in\Omega_{g}}n_{p}\mathcal{H}_{pp}+\sum\limits _{q,p\in\Omega_{g}}\Pi_{qp}\mathcal{K}_{pq}\,,\;\Omega{}_{g}\in\Omega_{\mathrm{II}}\label{Epair}
\end{equation}
with 
\begin{equation}
\Pi_{pq}=\left\{ \begin{array}{c}
\sqrt{n_{p}n_{q}}\qquad p=q\textrm{ or }p,q>\frac{N_{\mathrm{II}}}{2}\\
-\sqrt{n_{p}n_{q}}\qquad p=g\textrm{ or }q=g\qquad\;
\end{array}\right.
\end{equation}
is the energy of a pair of electrons with opposite spins. Eq. (\ref{Epair})
reduces to the NOF obtained from a ground-state singlet wavefunction,
so $E_{g}$ describes accurately two-electron systems \cite{Piris2018a}.
$\mathcal{K}_{pq}$ are the exchange integrals $\left\langle pq|qp\right\rangle $.

In the the last term of Eq. (\ref{EPNOF7}), $E_{fg}$ correlates
the motion of electrons with parallel and opposite spins belonging
to different subspaces ($\Omega_{f}\neq\Omega{}_{g}$): 
\begin{equation}
E_{fg}=\sum\limits _{p\in\Omega_{f}}\sum\limits _{q\in\Omega_{g}}\left[n_{q}n_{p}\left(2\mathcal{J}_{pq}-\mathcal{K}_{pq}\right)-\Phi_{q}\Phi_{p}\mathcal{K}_{pq}\right]\label{Efg}
\end{equation}
In Eq. (\ref{Efg}), $\mathcal{J}_{pq}$ are the usual Coulomb integrals
$\left\langle pq|pq\right\rangle $, whereas $\Phi_{p}=\sqrt{n_{p}(1-n_{p})}$.
It is not difficult to verify \cite{Piris2019} that the PNOF7 reconstruction
leads to $\mathrm{<}\hat{S}^{2}\mathrm{>}=S\left(S+1\right)$ with
$S=\mathrm{N_{I}}/2$. Therefore, the PNOF7 results should be exempt
from spin contamination effects.

Being an electron-pair-based functional \cite{Piris2018b}, PNOF7
is not capable of recovering the entire dynamic correlation, so we
must resort to perturbative corrections if we want to obtain significant
total energies. PNOF7 provides the NOs to form the reference energy
$\tilde{E}_{hf}$ in Eq. (\ref{Etotal}), namely,
\begin{equation}
\tilde{E}_{hf}=2\sum\limits _{g=1}^{\mathrm{N}_{\Omega}}\mathcal{H}_{gg}+\sum_{f,g=1}^{\mathrm{N}_{\Omega}}\left(2\mathcal{J}_{fg}-\mathcal{K}_{fg}\right)-\sum_{g=\frac{\mathrm{N_{II}}}{2}+1}^{\mathrm{N}_{\Omega}}\frac{\mathcal{J}_{gg}}{4}\label{HFL}
\end{equation}
In Eq. (\ref{HFL}), the last term eliminates the $\alpha\beta$-contribution
to the energy of the singly occupied orbitals since in each pure state
$\left|SM_{s}\right\rangle $ of the ensemble there is no such interaction.
In this sense, the zeroth-order Hamiltonian for the modified MP2 is
constructed from a closed-shell-like Fock operator that contains a
HF density matrix with doubly ($2n_{g}=2$) and singly ($2n_{g}=1$)
occupied orbitals.

$E^{sta}$ is the sum of the static intra-space and inter-space correlation
energies: 
\begin{equation}
\begin{array}{c}
E^{sta}=\sum\limits _{g=1}^{\mathrm{N_{II}/2}}\sum\limits _{q\neq p}\sqrt{\Lambda_{q}\Lambda_{p}}\,\Pi_{qp}\mathcal{\,K}_{pq}\\
\qquad-4\sum\limits _{f\neq g}^{\mathrm{N_{\Omega}}}\sum\limits _{p\in\Omega_{f}}\sum\limits _{q\in\Omega_{g}}\Phi_{q}^{2}\Phi_{p}^{2}\mathcal{K}_{pq}
\end{array}\label{Esta}
\end{equation}
where $\Lambda_{p}=1-\left|1-2n_{p}\right|$ is the amount of intra-space
static correlation in each orbital as a function of its occupancy.
Note that $\Lambda_{p}$ goes from zero for empty or fully occupied
orbitals to one if the orbital is half occupied.

$E^{dyn}$ is obtained from the second-order correction $E^{\left(2\right)}$
of the MP2 method. The first-order wavefunction is a linear combination
of all doubly excited configurations, considering one electron with
$\alpha$ or $\beta$ spin in $\Omega_{\mathrm{I}}$. The dynamic
energy correction takes the form 
\begin{equation}
E^{dyn}=\sum\limits _{g,f=1}^{\mathrm{N_{\Omega}}}\;\sum\limits _{p,q>\mathrm{N}_{\Omega}}^{N_{B}}A_{g}A_{f}\left\langle gf\right|\left.pq\right\rangle \left[2T_{pq}^{gf}\right.\left.-T_{pq}^{fg}\right]\label{E2}
\end{equation}
where 
\begin{equation}
\qquad\qquad A_{g}=\left\{ \begin{array}{c}
1,\quad\thinspace1\leq g\leq\mathrm{N_{II}}/2\\
\thinspace\thinspace\tfrac{1}{2},\quad\mathrm{N_{II}}/2<g\leq\mathrm{N}_{\Omega}
\end{array}\right.
\end{equation}
and $N_{B}$ is the number of basis functions. The amplitudes $T_{pq}^{fg}$
are obtained by solving the modified equations for the MP2 residuals
\cite{Piris2018c}. In order to avoid double counting of the electron
correlation, the amount of dynamic correlation in each orbital $p$
is defined by functions $C_{p}$ of its occupancy, namely, 
\begin{equation}
\begin{array}{c}
C_{p}^{tra}=\begin{cases}
\begin{array}{c}
\begin{array}{c}
1-4\left(1-n_{p}\right)^{2},\end{array}\\
1-4n_{p}^{2},
\end{array} & \begin{array}{c}
p\leq\mathrm{N}_{\Omega}\\
p>\mathrm{N}_{\Omega}
\end{array}\end{cases}\\
\:C_{p}^{ter}=\begin{cases}
\begin{array}{c}
\begin{array}{c}
1,\end{array}\\
1-4\left(1-n_{p}\right)n_{p},
\end{array} & \begin{array}{c}
p\leq\mathrm{N}_{\Omega}\\
p>\mathrm{N}_{\Omega}
\end{array}\end{cases}
\end{array}\label{Cp}
\end{equation}
where $C_{p}$ is divided into intra-space ($C_{p}^{tra}$) and inter-space
($C_{p}^{ter}$) contributions. According to Eq. (\ref{Cp}), fully
occupied and empty orbitals yield a maximal contribution to dynamic
correlation, whereas orbitals with half occupancies contribute nothing.
It is worth noting that $C_{p}^{ter}$ is not considered if the orbital
is below $\mathrm{N_{\Omega}}$. Using these functions as the case
may be (intra-space or inter-space), the modified off-diagonal elements
of the Fock matrix ($\tilde{\mathcal{F}}$) are defined as 
\begin{equation}
\tilde{\mathcal{F}}_{pq}=\begin{cases}
C_{p}^{tra}C_{q}^{tra}\mathcal{F}_{pq}, & p,q\in\Omega_{g}\\
C_{p}^{ter}C_{q}^{ter}\mathcal{F}_{pq}, & otherwise
\end{cases}
\end{equation}
as well as modified two-electron integrals: 
\begin{equation}
\widetilde{\left\langle pq\right|\left.rt\right\rangle }=\begin{cases}
C_{p}^{tra}C_{q}^{tra}C_{r}^{tra}C_{t}^{tra}\left\langle pq\right|\left.rt\right\rangle , & p,q,r,t\in\Omega_{g}\\
C_{p}^{ter}C_{q}^{ter}C_{r}^{ter}C_{t}^{ter}\left\langle pq\right|\left.rt\right\rangle , & otherwise
\end{cases}
\end{equation}
where the subspace index $g=1,...,\mathrm{N}_{\Omega}$. This leads
to the following linear equation for the modified MP2 residuals: 
\begin{equation}
\widetilde{\left\langle ab\right|\left.ij\right\rangle }+\left(\mathcal{F}_{aa}\right.+\mathcal{F}_{bb}-\mathcal{F}_{ii}-\left.\mathcal{F}_{jj}\right)T_{ab}^{ij}\:+\label{residual}
\end{equation}
\[
{\displaystyle \sum_{c\neq a}\mathcal{\tilde{F}}_{ac}T_{cb}^{ij}}+{\displaystyle \sum_{c\neq b}}T_{ac}^{ij}\mathcal{\tilde{F}}_{cb}-{\displaystyle \sum_{k\neq i}}\tilde{\mathcal{F}}_{ik}T_{ab}^{kj}-{\displaystyle \sum_{k\neq j}}T_{ab}^{ik}\mathcal{\tilde{F}}_{kj}=0
\]
where $i,j,k$ refer to the strong occupied NOs, and $a,b,c$ to weak
occupied ones. It should be noted that diagonal elements of the Fock
matrix ($\mathcal{F}$) are not modified. By solving this linear system
of equations the amplitudes $T_{pq}^{fg}$ are obtained, which are
inserted into the Eq. (\ref{E2}) to achieve $E^{dyn}$.

\section{Results and Discussion}

The appropriateness of the selected splitting of the NOs space, $\Omega$,
sketched in Fig. \ref{fig1}, is confirmed by the data shown in Table
\ref{tab:s_t}. The PNOF7-MP2 calculated singlet/triplet splitting
energy of the Sc$^{+}$ cation compares well with its corresponding
experimental mark. It is worth noting that MCQDPT result is very satisfactory,
``for the right reason'', and the MP4 one is even better, but, for
``the wrong reason'', given the large multiconfigurational character
of the electronic structure of both the ground $^{3}{\rm D}$ and
the first excited singlet $^{1}{\rm D}$ states of Sc$^{+}$. The
performance of PNOF7-MP2 for the dissociation energy of the {[}Sc(OH)$_{2}${]}$^{+}$
ion-molecule complex is particularly satisfactory, for it lies within
the error-bar range of the experimental mark. Observe that MP4 performs
erratically for the prediction of this dissociation energy. For the
scandium oxide cation, PNOF7-MP2 predicts a singlet/triplet splitting
energy slightly larger than MCQDPT. Having no available experimental
results for this species, the similarity of PNOF7-MP2 and MCQDPT values
suggest that PNOF7-MP2 should be seen as an accurate estimation. Recall
that once again, MP4 does wildly wrong for this particular singlet/triplet
splitting energy.

\begin{table}[t]
\centering{}\caption{\label{tab:s_t}Singlet-Triplet energy splittings, $\Delta_{S/T}$,
in eV, for the Sc$^{+}$ and for ScO$^{+}$, and dissociation energy,
D$_{0}$, in eV, for the {[}Sc(OH)$_{2}${]}$^{+}$ ion-molecule complex.
Zero point vibrational energy corrections have been included.}
\begin{tabular}{l|lcc}
 & $\Delta_{S/T}^{({\rm Sc}^{+})}$ & $\Delta_{S/T}^{({\rm ScO}^{+})}$ & ${\rm D}_{0}$\tabularnewline
\hline 
\hline 
NOF-MP2 & 0.475 & -3.762 & 1.264\tabularnewline
MCQDPT & 0.380 & -3.376 & 1.456\tabularnewline
CCSDT/TZVP+\cite{aran1999} & 0.550 & -3.496 & 1.410\tabularnewline
B3LYP/TZVP+\cite{aran1999} & 0.916 & -3.292 & 1.580\tabularnewline
MCSCF \cite{tilson1991} &  & -3.450 & 1.571\tabularnewline
MP4 \cite{ye1997} & 0.300 & -5.229 & 2.500\tabularnewline
Exp. & 0.315\cite{sugar:1985} &  & 1.36$\pm$0.13\cite{magnera1989}\tabularnewline
\hline 
\end{tabular}
\end{table}

The energetics of the dehydrogenation of water by the Sc$^{+}$ cation,
reactions (\ref{rxn1}) and (\ref{rxn2}), has also been established
with a high degree of confidence. The experimental data available
for reaction (\ref{rxn1}), and the highly accurate data available
for both reactions, has been collected in Table \ref{tab1}. 
\begin{eqnarray}
\mathrm{Sc}^{+}(^{3}\mathrm{D})+\mathrm{H_{2}O} & \rightarrow & \mathrm{ScO}^{+}(^{1}\mathrm{\Sigma})+\mathrm{H_{2}}+\Delta{\rm E}_{1}\label{rxn1}\\
\mathrm{Sc}^{+}(^{3}\mathrm{D})+\mathrm{H_{2}O} & \rightarrow & \mathrm{ScO}^{+}(^{3}\mathrm{\Delta})+\mathrm{H_{2}}+\Delta{\rm E}_{2}\label{rxn2}
\end{eqnarray}

Armentrout et al. \cite{Aristov1984}, reported in 1984 a value of
1.866$\pm$0.304 eV for reaction energy of reaction (\ref{rxn1}),
and ten years later they refined their experimental uncertainty by
on order of magnitude, given an experimental estimate of 2.03$\pm$0.06
eV \cite{chen1994}. All the calculated reactions energies for reaction
(\ref{rxn1}), shown in Table \ref{tab1}, except MP4, lie below the
lower bound of the best experimental reaction energy value. However,
it is worth noting that PNOF7-MP2 gets the closer of them all, namely,
only 6 meV off the mark. The erratic behavior of MP4 calculations
for this kind of reactions is manifested here by the calculated overwhelmingly
large reaction energy of 4.687 eV. Recall that both B3LYP and CCSD(T)
calculations lead to predicted reaction energies in pretty close agreement
with both the PNOF7-MP2 and MCQDPT values and the experimental estimate.
MCSCF, which lacks an explicit consideration of the dynamical electron
correlation effects, yields a reaction energy $\sim$0.5 eV lower
than methods that treat explicitly dynamical correlation effects.
This emphasizes the fact that the dynamical electron correlation effects
must be accounted for, in addition to the static electron correlation
effects, in order to yield reliable reaction energies.

\begin{table}[t]
\centering{}\caption{\label{tab1}$\Delta$E$_{1}$ is the energy in eV, including the
zero-point vibrational energy corrections, for reaction (\ref{rxn1}),
and $\Delta$E$_{2}$ for reaction (\ref{rxn2}).}
\begin{tabular}{l|cc}
 & $\Delta$E$_{1}$ & $\Delta$E$_{2}$\tabularnewline
\hline 
\hline 
NOF-MP2 & 1.964 & -1.323\tabularnewline
MCQDPT & 1.972 & -1.404\tabularnewline
CCSDT/TZVP+\cite{aran1999} & 1.956 & -1.518\tabularnewline
B3LYP/TZVP+\cite{aran1999} & 1.939 & -1.253\tabularnewline
MCSCF\cite{tilson1991} & 1.415 & -2.034\tabularnewline
MP4 \cite{ye1997} & 4.687 & \tabularnewline
Exp.\cite{chen1994} & 2.03$\pm$0.06 & \tabularnewline
Exp.\cite{Aristov1984} & 1.866$\pm$0.304 & \tabularnewline
\hline 
\end{tabular}
\end{table}

\begin{figure}
\centering{}\includegraphics[width=1\linewidth]{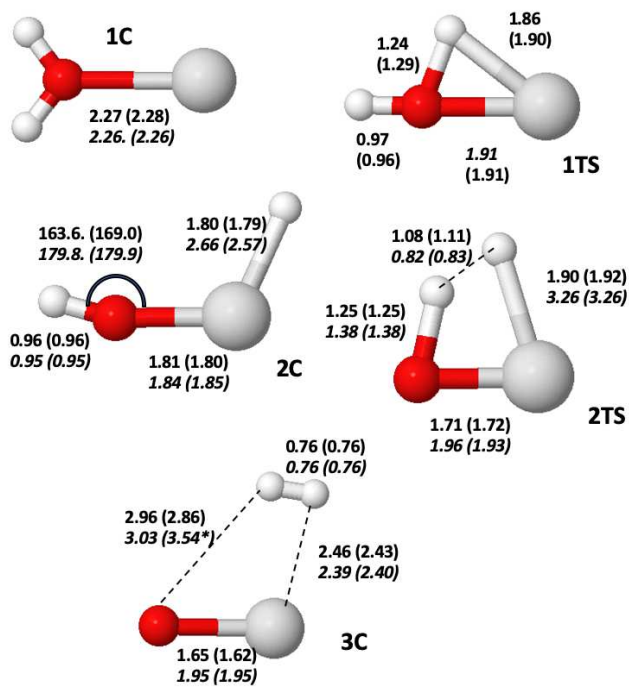} \caption{Optimized geometries of intermediate stationary points of the singlet
and triplet potential energy surfaces. Plain text values correspond
to the singlet optimum geometries; PNOF7 and, MCSCF in parenthesis.
Values in italics correspond to their corresponding triplet optimum
geometries. The 3C optimum MCSCF geometry, has the H$_{2}$ moiety
rotated with respect to the PNOF7 geometry. Thus, the MCSCF H--O
distance marked with the \textquotedblleft$\ast$\textquotedblright ,
turns out to be sizeable larger than its corresponding PNOF7 distance.
Images were created with Jmol \cite{jmol}.}
\label{fig:figGeom}
\end{figure}

The reliability of the optimized geometrical structures of all the
intermediates of reactions (\ref{rxn1}) and (\ref{rxn2}), is addressed
next. The structure of the first intermediate, the encounter ion-molecule
complex, {[}Sc(OH$_{2}$){]}$^{+}$ (see structure 1C of Fig. \ref{fig:figGeom}),
has recently been investigated by inspecting the O--H stretching
region using infrared laser photodissociation and the method of rare
gas atom predissociation. The measured O--H stretches have been found
to be shifted to lower frequencies than those for the free water molecule.
The significant experimental data, along with our calculated values
can be found in Table \ref{tab:freq}.

\begin{table}[ht]
\centering{}\caption{\label{tab:freq}OH symmetric and antisymmetric stretching vibrational
frequencies, in cm$^{-1}$, and red shifts (in parentheses) with respect
to isolated water vibrational bands for the {[}Sc(OH$_{2}$){]}$^{+}$
ion-molecule complex. The experimental values for the ion-molecule
complex, which have an argon tag atom bound on the metal cation, have
been/ taken from Ref \cite{Carnegie2011}. The reference experimental
frequencies for OH$_{2}$ have been taken from \cite{Shimanouchi},
and the theoretical ones have been calculated at the same level level
of theory using the harmonic approximation.}
\begin{tabular}{c|ccc}
 & Exp. & PNOF7 & MCSCF(10,17)\tabularnewline
\hline 
\hline 
$\nu_{symm}$ & 3580 (77) & 3730 (61) & 3746 (24)\tabularnewline
$\nu_{asymm}$ & 3656 (100) & 3811 (45) & 3821 (67)\tabularnewline
\end{tabular}
\end{table}

The experimentally measured frequencies and the red shits of both,
the O--H symmetric and anti-symmetric stretching vibrational bands
are nicely predicted by our PNOF7 calculations, as shown in Table
\ref{tab:freq}. The discrepancies of the calculated red shifts with
respect to their corresponding experimental values, can reasonably
be ascribed to the Argon tag-atom attached to the scandium in the
experiments, and to the harmonic approximation used in the theoretical
calculations. In spited of all this, on the whole, the agreement between
theory and experiment is remarkably good.

The electronic structure of the {[}Sc(OH$_{2}$){]}$^{+}$ ion-mo-lecule
complex deserves a further comment. Indeed, as revealed by the inspection
of the ONs of both the MCSCF and the PNOF7 calculations, shown the
Table \ref{tab:spin-flip}, its singlet state corresponds to an almost
pure open-shell configuration. The corresponding minimum energy isomer
triplet spin state results from an spin-flip, having rather the same
ONs, but with a ferromagnetic like coupling of the spins of the two
singly occupied NOs. Clearly, both PNOF7 and MCSCF closely agree on
this prediction. A feature that can hardly be captured by single-configuration
type methods, including most current approximate DFT implementations.

\begin{table}[ht]
\centering{}\caption{\label{tab:spin-flip}Natural occupations numbers of the singlet and
triplet spin states of the {[}Sc(OH$_{2}$){]}$^{+}$ ion-molecule
complex, at the MCSCF and PNOF7 levels of theory. MCSCF orbitals 8--11
belong the fixed occupation inactive core orbitals' set.}
\begin{tabular}{c|cc|cc}
 & \multicolumn{2}{c|}{Singlet 1C} & \multicolumn{2}{c}{Triplet 1C}\tabularnewline
\hline 
\hline 
Orb. \# & MCSCF & NOF & MCSCF & NOF\tabularnewline
\hline 
08 & 2.00 & 1.99 & 2.00 & 1.99\tabularnewline
09 & 2.00 & 1.98 & 2.00 & 1.98\tabularnewline
10 & 2.00 & 1.98 & 2.00 & 1.98\tabularnewline
11 & 1.98 & 1.98 & 1.98 & 1.98\tabularnewline
12 & 1.97 & 1.98 & 1.97 & 1.98\tabularnewline
13 & 1.97 & 1.98 & 1.97 & 1.99\tabularnewline
14 & 1.96 & 1.97 & 1.96 & 1.97\tabularnewline
15 & 1.01 & 1.16 & 1.00 & 1.00\tabularnewline
16 & 0.91 & 0.81 & 1.00 & 1.00\tabularnewline
\hline 
\end{tabular}
\end{table}

The singlet and triplet potential energy surfaces for the dehydrogenation
of water by Sc$^{+}$ has been profusely investigated. Thus, Tilson
and Harrison \cite{tilson1991} carried an exhaustive MCSCF+1+2 investigation
with a relatively small basis set. Later, Ye carried out single reference
MP2 like calculations, and finally Irigoras et al. \cite{aran1999},
and Russo et al. \cite{nino} made a thorough scan of both potential
energy surfaces using the hybrid B3LYP exchange-correlation functional
complemented with CCSD(T) single point calculations for the refinement
of the energies. All these investigations yield the same qualitative
picture. Thus, in order to assess the reliability of our PNOF7/TZVP+
optimum intermediates' structures we shall compare them with the results
of MCSCF(10,17) geometry optimization carried out with the TZVP+ basis
set. The salient geometrical features of the optimized structures
are displayed in Fig. \ref{fig:figGeom}.

The optimum PNOF7 and MCSCF geometries for the five intermediates
shown in Fig. \ref{fig:figGeom} are very similar. Likewise, they
are also similar to the B3LYP optimum geometries reported earlier
by Irigoras et al. \cite{aran1999} and Sicilia et al. \cite{nino}.
The only remarkable difference is that for the triplet-spin potential
energy surface the exit channel ScO$^{+}\cdots$ H$_{2}$ complex
is planar, while the MCSCF favors the H$_{2}$ fragment rotated 90
degrees out of the plane with respect to the ScO$^{+}$ molecular
axis.

\begin{figure*}
\centering{}\includegraphics[scale=0.5]{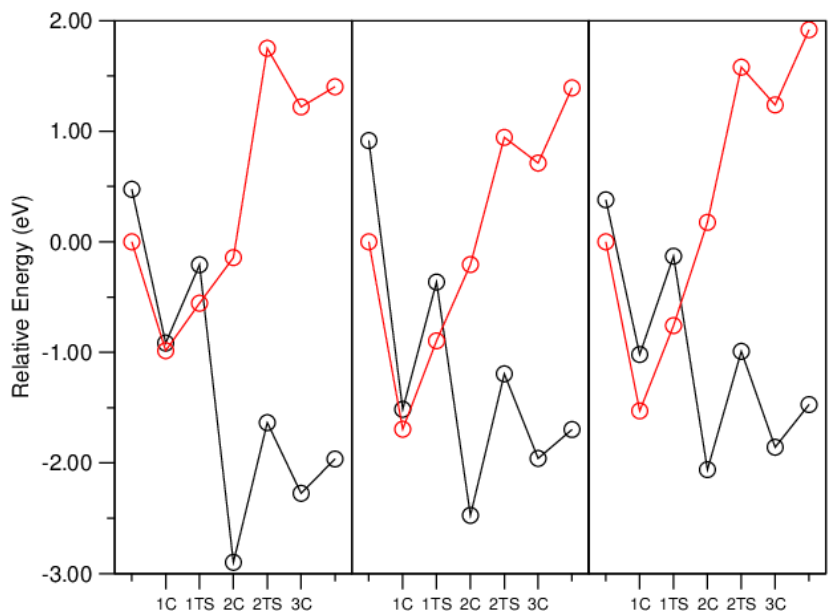} \caption{Potential energy surfaces for the triplet spin state (red curves)
and for the singlet spin (black curves) state, following the Sc$^{+}$+OH$_{2}\rightarrow$ScO$^{+}$+H$_{2}$
reaction path. Left panel: PNOF7-MP2, center panel: B3LYP, right panel:
MCQDPT.}
\label{pes}
\end{figure*}

Finally, Fig. \ref{pes} shows the schematic reaction paths for the
Sc$^{+}$+OH$_{2}\rightarrow$ScO$^{+}$+H$_{2}$ reaction on the
singlet and triplet potential energy surfaces, evaluated at the PNOF7-MP2,
B3LYP and MCQDPT levels of theory with the TZVP+ basis set. Although
the relative energies of the various intermediates of the proposed
mechanism vary slightly among the levels of theory used, the chemistry
coming out from the inspection of the three panels of Fig. \ref{pes}
is the same. Namely, the only exothermic product is ScO$^{+}$($^{1}\mathrm{\Sigma}$)
+ H$_{2}$ ($^{1}\mathrm{\Sigma}_{g}^{+}$), which can be reached
starting either from the singlet spin reactants, Sc$^{+}$($^{1}\mathrm{D}$)
+ OH$_{2}$($^{1}\mathrm{A}_{1}$), or from the triplet spin state
ones, Sc$^{+}$($^{3}\mathrm{D}$) + OH$_{2}$($^{1}\mathrm{A}_{1}$),
by virtue of the spin crossing occurring in the region between the
1TS and 2C intermediates.

\section{Conclusions}

Current formulations of the natural orbital functional theory based
on the Piris ansatz have yield a family of energy functionals of the
1RDM which treat accurately static electron correlation effects, and
a substantial portion of the dynamic electron correlation effects
\cite{Mitxelena2019}. The missing part of the dynamical electron
correlation can be brought back in by supplementing the natural orbital
functional theory calculations with a perturbational scheme. The implementation
of such a scheme into a practical user-friendly code for molecular
electronic structure calculations has recently been achieved \cite{Piris2021}.
This scheme has the advantage to treating electron correlation effects
better than current implementations of DFT, and the disadvantage of
being more demanding computationally than DFT. Yet its is far less
demanding than multiconfigurational theory wave function type implementations.
As an example, MCQDPT triplet calculations required around 3 days,
300GB of RAM and more than 20TB of storage, while NOF-MP2 calculations
lasted less than 10 hours and required less that 8 GB of RAM.

Herein, we have shown that the recently proposed PNOF7 supplemented
with second-order Moller-Plesset calculations, PNOF7-MP2, is very
reliable for accurate chemical reaction mechanistic studies of elementary
reactions of transition metal compounds, for which strong electron
correlation effects are known to be ubiquitous. We have investigated
the dehydrogenation of water by the scandium cation and found that
(i) the singlet-triplet energy gaps of Sc$^{+}$, and ScO$^{+}$ cations
are accurately reproduced, (ii) the dissociation energy of the {[}Sc(OH$_{2}$){]}$^{+}$
ion-molecule complex is also estimated within the experimental error
bars, (iii) the calculated frequencies and red shifts of the OH symmetric
and anti-symmetric vibrational bands agree satisfactorily with experimental
measurements, (iv) the open-shell electronic structure of the singlet
spin state of {[}Sc(OH$_{2}$){]}$^{+}$ is also correctly described,
along with that of its spin-flip related ground state triplet spin
state, (v) the calculated geometries of the intermediate species on
both, the singlet and triplet state potential energy surfaces compare
satisfactorily with those obtained at the MCSCF level of theory, and
(vi) the overall energetics of both potential energy surfaces come
in close agreement with those of the highly accurate MCQDPT/MCSCF
calculations.

All in all, we believe that PNOF7-MP2 deserves serious consideration
as a reliable alternative for chemical reaction mechanistic studies
where strong electron correlation effects play a role.
\begin{acknowledgments}
The authors thank for technical and human support provided by IZO-SGI
SGIker of UPV/EHU and European funding (ERDF and ESF) and DIPC for
the generous allocation of computational resources. Financial support
comes from the Spanish Office for Scientific Research (MCIU /AEI /FEDER,
UE), Ref.:PGC2018-097529-B-100 and Eusko Jaurlaritza (Basque Government),
Ref.: IT1254-19.
\end{acknowledgments}


\end{document}